# A study of thinner-THGEM, with some applications

H. B. Liu[a,b*], Q. Liu[a], S. Chen[a,b], X. D. Ruan[b], C. Nicholson[a], Y. G. Xie[a,c], Y. H. Zheng[a], Z. P. Zheng[c], J. G. Lu[c], L. Zhou[c], A. S. Tang[d], Y. D. Yang[d], Y. Dong[d], and M. Li[d]

[a] *Graduate University of Chinese Academy of Sciences,*
  *100049, Beijing, China*

[b] *Guangxi University,*
  *530004, Nanning, China,*

[c] *Institute of High Energy Physics, Chinese Academy of Sciences,*
  *100049, Beijing, China*

[d] *Second Academy of China Aerospace Science and Industry CORP.,*
  *100039, Beijing, China*

  *E-mail:* `liuhb@ihep.ac.cn`

ABSTRACT: THGEMs (THick Gas Electron Multiplier) of varying thickness, hole diameter and hole pitch have been studied. For a thinner-THGEM of thickness 0.2 mm, with hole diameter 0.2 mm, pitch 0.2 mm and narrow (5-10 μm) rim, the performance of gain versus high voltage with different gas mixtures have been studied using 5.9 keV X-rays. In general, a gain of around $3 \times 10^3$ was obtained with a single board in $Ar/iC_4H_{10}$, and gains higher than $10^4$ were obtained in Ne mixture at lower voltage. The dependence of the energy resolution on the drift and induction electric fields was measured and an energy resolution of 15.9% was obtained. A curved thinner-THGEM chamber with one-dimensional readout has been assembled for use in the diffraction studies at the Beijing Synchrotron Radiation Facility (BSRF). A cosmic-ray muon hodoscope based on thinner-THGEM was developed as a teaching experiment for the Graduate University of the Chinese Academy of Sciences (GUCAS).

---

[*] Corresponding author.

# Contents

## 1. Introduction to the THGEM at GUCAS and IHEP

The gas electron multiplier (GEM) was invented by F. Sauli at CERN in 1997 [1]. Thicker GEM (THGEM), a geometrical extension of standard GEMs, was developed by A. Breskin and others in 2004 [2]-[4]. THGEM is robust, cheap and can easily be manufactured using standard printed circuit board (PCB) techniques.

More than eighteen types of THGEMs have been produced and studied at the Graduate University of the Chinese Academy of Sciences (GUCAS) and the Institute for High Energy Physics (IHEP). Two sets of THGEM of 0.5 mm and 1.0 mm thickness respectively were designed and built for studying the dependence of gas gain on different hole diameters. The schematic layouts of these THGEMs are shown in Figure 1. Each THGEM had four sectors, each sector having different hole diameters as follows: 300 µm (b), 400 µm (a), 500 µm (d), 600 µm (c). All holes on the four sectors had the same 1 mm pitch. Each sector had the same active area of $3 \times 4$ cm$^2$.

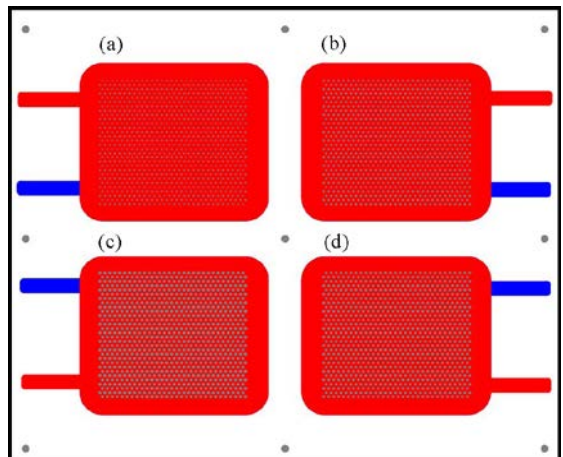

**Figure 1:** Schematic diagram of the THGEM with four sectors of different hole diameters: (a) 400µm, (b) 300 µm, (c) 600 µm, (d) 500 µm.

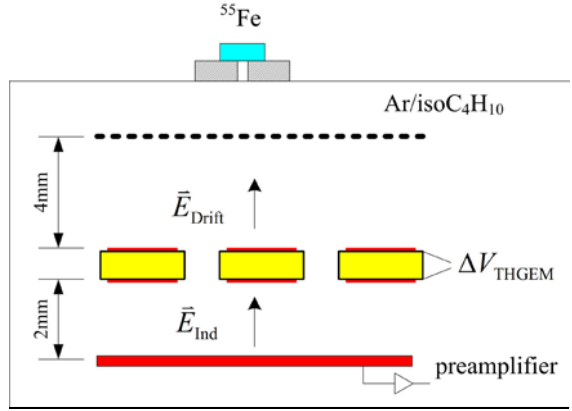

**Figure 2:** Schematic diagram of the THGEM detector.

Figure 2 provides a schematic view of the detector configuration, consisting of a cathode and anode with a single THGEM in between.

Irradiated with a $^{55}$Fe X-ray source, the gas of the THGEM detector is ionized. The ionized electrons in the upper drift region are drifted toward the THGEM and multiplied in the THGEM holes, and then enter the lower inductive region. Finally, the electrons are collected by the anode. All electrodes were biased with CAEN SY127 HV power supplies and the signal was recorded using an ORTEC 142 preamplifier (charge sensitivity 1V/pC) followed by an ORTEC 450 amplifier (shaping time $\tau$=0.5 μs) and an ORTEC 916 multi-channel analyzer (MCA). The THGEM was mounted on a polyethylene (PE) frame and installed in a gas-tight epoxy-made chamber, which was continuously flushed with 1 atm of pre-mixed gas $Ar/CO_2/CH_4$(89/10/1).

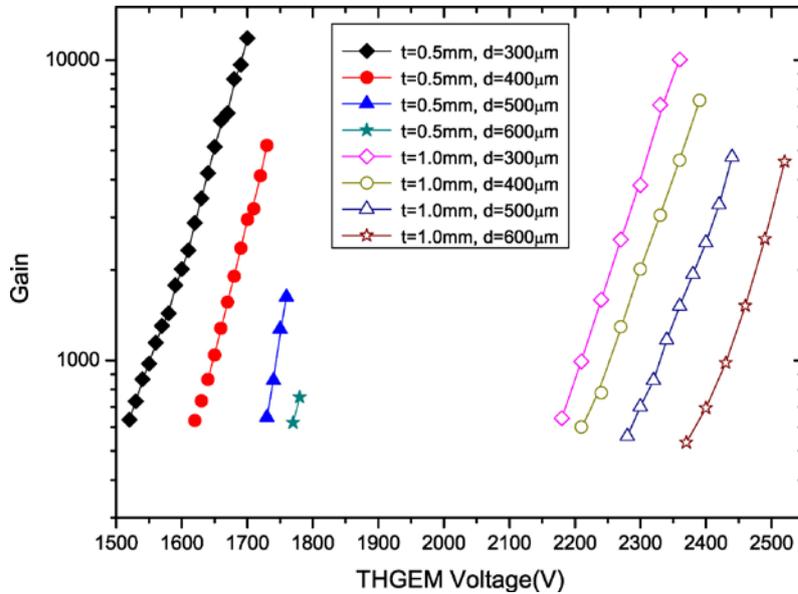

**Figure 3:** Gain as a function of the THGEM voltage $\Delta V_{THGEM}$, measured with 5.9 keV X-rays in different configurations [THGEM thickness t=0.5 mm (solid symbol), t=1.0 mm (open symbol) and THGEM hole diameter d=300 μm (rhombus), d=400 μm (circle), d= 500μm (triangle), d=600 μm (star)]. We define the maximum applied voltage as the one where the discharge rate reaches the rate of one every

30 seconds.

After calibration, the gains for eight kinds of THGEM were obtained. The results in Figure 3 show that the thinner the thickness the lower the working voltage is applied for the same gas gain; also, for thinner THGEM, higher gain is obtained for smaller hole diameters.

When electrons are avalanching under high electric field in THGEM holes, discharge is mainly triggered by the imperfections of THGEM, such as copper burr around the rim, dielectric polarization of FR4 substrate and hole imperfection, etc. The THGEM with small hole size can operates at relatively low voltage and thus decreases the possibility of discharge.

## 2. The performance of Thinner-THGEM

A type of thinner-THGEM was developed at GUCAS and IHEP with a thickness of 0.2 mm, hole diameter of 0.2 mm, pitch of 0.5 mm and a very small rim of 5~10 μm.

Figure 4 shows the gain curves of the thinner-THGEM for different gas mixtures. All curves were obtained with the same detector set-up and collimated $^{55}$Fe X-ray source shown in Figure 2. As expected, for the same gain, the thinner-THGEM working voltage is lower than the working voltage (as obtained in previous works [3]-[5]) for ordinary THGEM. The gains of thinner-THGEM were up to $3\times10^3$ with a single board in Ar/iC$_4$H$_{10}$ and higher than $10^4$ in Ne mixture at lower voltage (same as the results in reference [6]). For Ar/iC$_4$H$_{10}$, the lower the iC$_4$H$_{10}$ concentration the lower the operating voltage is applied for the same gas gain, as shown in Figure 4.

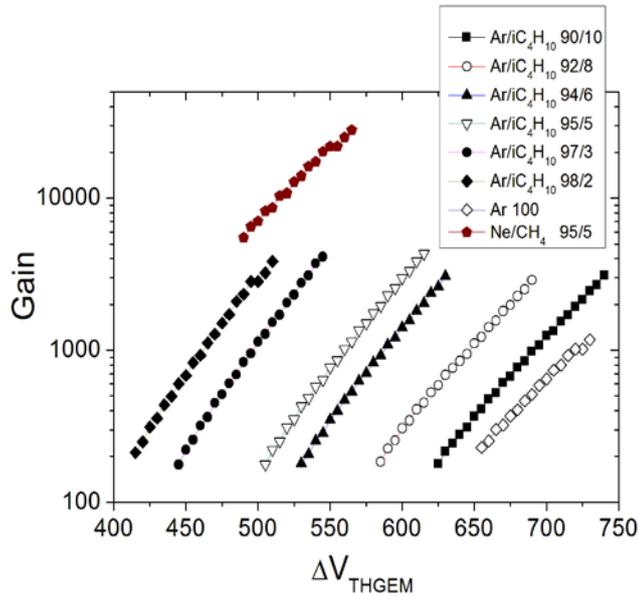

**Figure 4:** Gain curves in different gas mixtures. Very high gain is observed using the Neon mixture at relatively low operating voltages.

The stability test result is shown in Figure 5. The detector was operated at $\Delta V_{THGEM}$ =560V and illuminated by a collimated $^{55}$Fe X-ray source with a counting rate of ~400 X-ray

photons/mm$^2 \cdot$s in Ar/iC$_4$H$_{10}$ (95/5). The thinner-THGEM provided a stable gain from the very beginning.

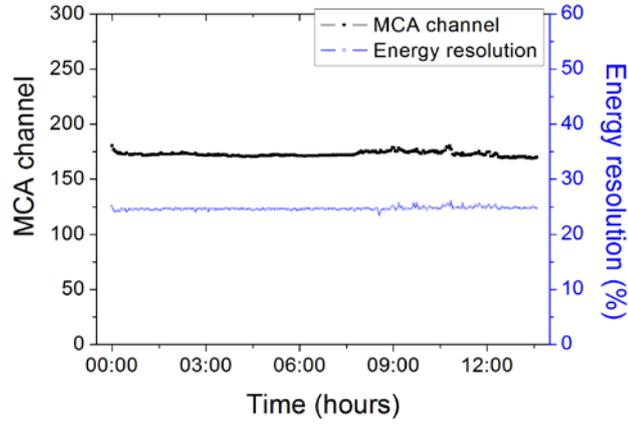

**Figure 5:** Fourteen hours test of the gain stability (upper line) and energy resolution (lower line) for thinner-THGEM at $\Delta V_{THGEM}$ =560V and a gain of $1.7\times10^3$.

With Ar/iC$_4$H$_{10}$(97/3), an energy resolution (FWHM) of 15.9% was obtained, as shown in Figure 6. In order to evaluate the influence of electric field in the drift region ($E_{drift}$) and electric field in the induction region ($E_{ind}$) on the energy resolution and the gas gain, $\Delta V_{THGEM}$ was set to a constant value of 560V. The energy spectrum of $^{55}$Fe was acquired by MCA and analyzed using ROOT [7].

The dependence of energy resolution on the drift electric field ($E_{drift}$) is shown in Figure 7. The electric fields $E_{drift}$ and $E_{ind}$ in the detector cover a range from 0.0 kV/cm to 2.5 kV/cm and 1.0 kV/cm to 4.0 kV/cm respectively.

Figure 7 shows that the value of $E_{drift}$ plays an important role in the THGEM detector performance, while $E_{ind}$ has almost no influence on the energy resolution (without considering the time resolution and spatial resolution in this measurement). A properly tuned $E_{drift}$ increases the probability of the primary ionization electrons to drift into the THGEM holes, thus resulting in good detector performance.

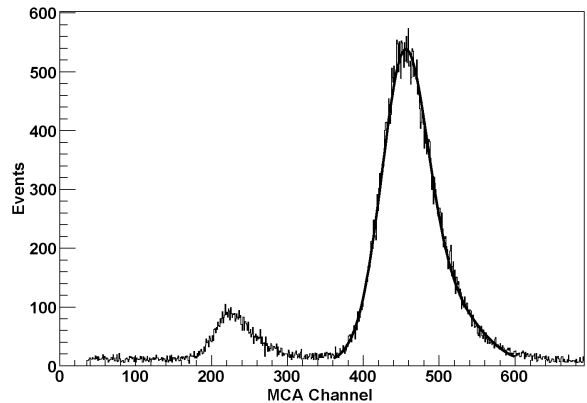

**Figure 6:** Energy resolution of 15.9% (FWHM) in $Ar/iC_4H_{10}$ (97/3) at $E_{drift}$=0.6 kV/cm, $E_{ind}$=4.5 kV/cm, $\Delta V_{THGEM}$=540 V and a gain of $3.7\times10^3$.

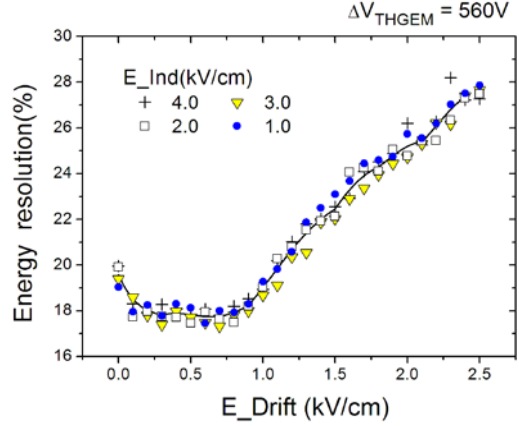

**Figure 7:** Dependence of the energy resolution (FWHM) on $E_{drift}$ under varying $E_{ind}$. The energy resolution depends strongly on $E_{drift}$ but has almost no dependence on $E_{ind}$. The optimum energy resolution was observed at $E_{drift}$=0.2~0.7 kV/cm in $Ar/iC_4H_{10}$ (95/5).

Compared to the THGEMs in our previous study [5], which had thickness of 0.4-1.0 mm, the thinner-THGEM gives better energy resolution, stability and fewer observed discharges at moderate gain operation.

## 3. Applications of thinner-THGEM

### 3.1 Applications for X-ray diffraction

A curved chamber with 1-D readout was made to provide a parallax-error-free [8] X-ray diffraction spectrometer for station 1W1A of the Beijing Synchrotron Radiation Facility (BSRF). The X-ray energy of the 1W1A station is 8.05 keV and 13.5 keV, and the 8.05 keV X-ray intensity is higher than $10^{11}$ photons/s. Based on the time structure and duty cycle (~200 ps beam width and 2 ns period) of BSRF, the expected results only depend on the accumulation effect. In this measurement, accumulated anode current was measured for the high intensity X-ray.

Figure 8 shows the curved thinner-THGEM and chamber with electric readout. The thinner-THGEM with 0.2 mm thickness can easily be curved to the desired shape. The distance between sample and the curved thinner-THGEM detector is 20 cm. The design parameters of the detector are: X-ray diffraction detection angle of 48°, angular resolution of 0.2°, 512 anode readout strips with pitch of 0.5 mm and width of 0.4 mm.

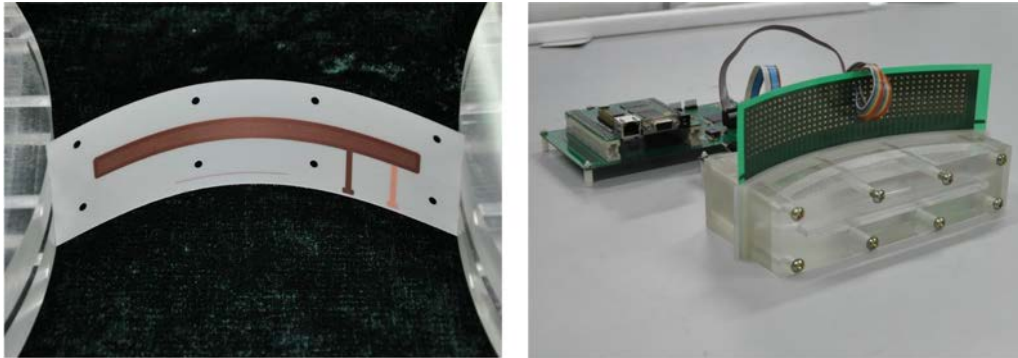

**Figure 8:** Curved thinner-THGEM (left) as an X-ray diffraction spectrometer. Current of the anode strip was readout form the top side of the chamber (right).

The detector was built according to these parameters but with only 32 readout strips in the initial configuration.

Figure 9 shows the schematic view of the diffraction spectrometer detector and Figure 10 shows the test setup. The detector was tested using 17 keV X-rays (Mo target) with a 0.5 mm slit collimator. The operating parameters for the single thinner-THGEM were $E_{drift}$=0.7 kV/cm, $\Delta V_{THGEM}$ =400 V, $E_{ind}$=3.0 kV/cm in Ne/CH$_4$(95/5). Charge from the anode strip was collected on the integration capacitor of the DDC232 (32-channel current-input analog-to-digital converter) and converted to a digital signal. The digital signal was readout by single chip micro-computer (SCM) via FPGA and output to the personal computer (PC). The PC was used for online display and data analysis by means of the LabVIEW package.

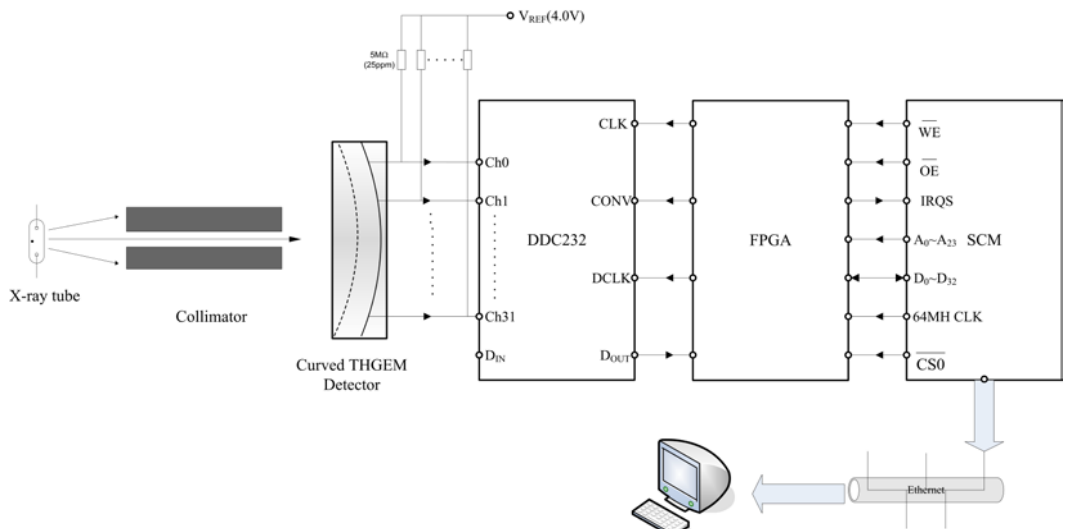

**Figure 9:** The schematic view of the diffraction spectrometer detector configuration. The DAQ system consisted of a 32-channel current-input analog-to-digital converter (DDC232), FPGA, single chip micro-computer (SCM) and a PC used for online display and data analysis.

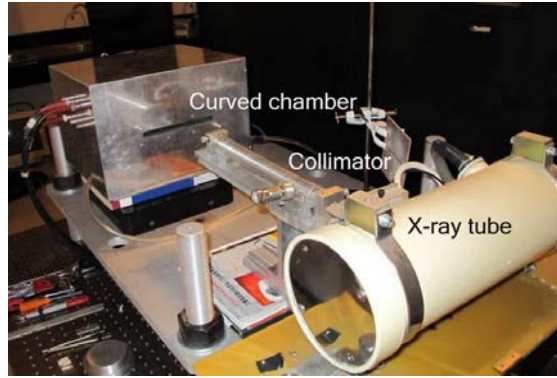

**Figure 10:** Test setup of the curved thinner-THGEM X-ray diffraction spectrometer.

As shown in Figure 11, a Gaussian fit of the current distribution along the strips (slit width 0.5 mm) gives σ of 1.4 strips and hence spatial resolution of ~0.7 mm.

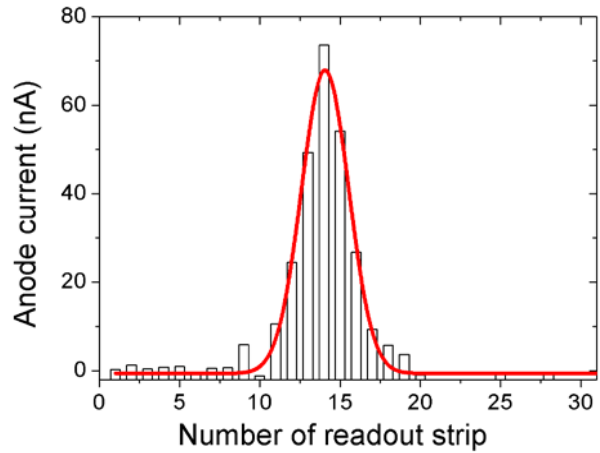

**Figure 11:** Current distribution along strips (slit width 0.5 mm). Spatial resolution is about 0.7 mm.

### 3.2 Muon hodoscope

A cosmic-ray muon hodoscope based on thinner-THGEM was built for a teaching experiment at GUCAS. The hodoscope consists of an upper and lower detector. Each detector has two cascaded thinner-THGEM boards with an effective area of 50×50 mm$^2$. The effective gain is related to the appropriate voltages applied to the THGEM, based on our previous studies. The electric fields of the detector were selected as follows: the drift electric field ($E_{drift}$) of 0.35 kV/cm, the transfer electric field ($E_{tran}$) of 3 kV/cm, the induction electric field ($E_{ind}$) of 3.3 kV/cm, and the voltage of both cascaded thinner-THGEMs ($\Delta V_{THGEM}$) of 500 V, giving a gas gain of the detector 3×10$^4$ in Ar/iC$_4$H$_{10}$ (95/3).

The readout anode consists of 16 pads, each pad with an area of 12.4×12.4 mm$^2$ and pitch of 12.6 mm, connected to the preamplifier via a short coaxial cable.

As shown in Figure 12, the DAQ system consists of two 16-channel fast preamplifiers (FBPANIK-04) with voltage amplification of 100, ALEPH muon detector electronics serial readout SGS (System for Gas Streamer tubes) card, single chip micro-computer (SCM), and a personal computer.

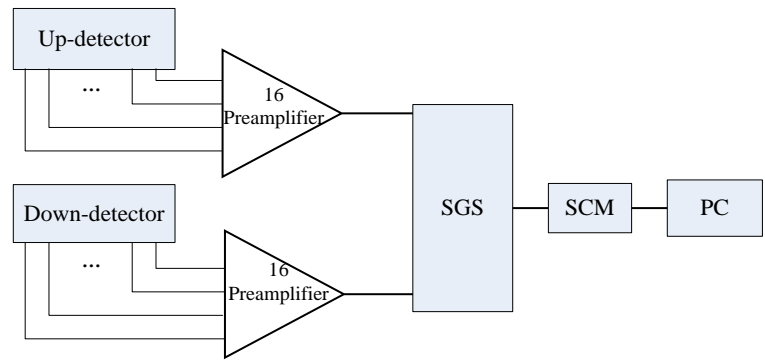

**Figure 12.** Hodoscope system for detecting cosmic ray muons.

The data acquisition and processing software were written using KEIL-C++ software for SCM and LabVIEW for online display and data analysis. The display screen was designed as follows: when any of the 32 pads, either the upper 16 or lower 16, is hit by a cosmic ray muon, a spot is displayed. When a muon passes through both upper and lower detectors simultaneously, a track line is displayed through these two pads on the computer screen, as shown in Figure 13.

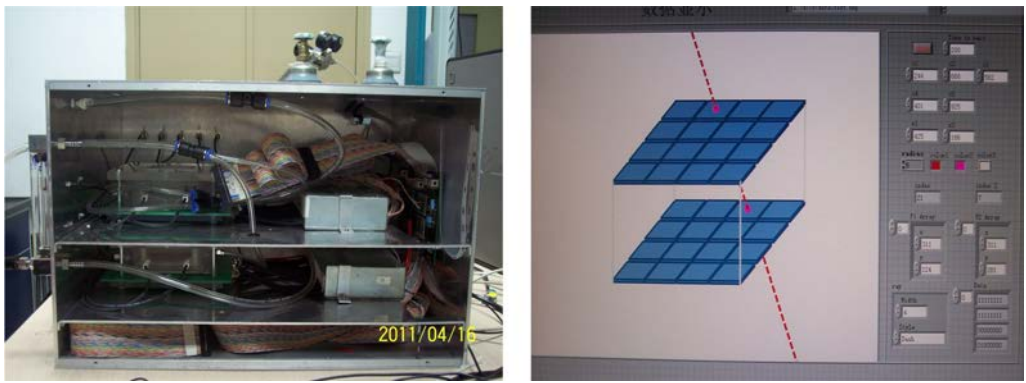

**Figure 13:** The muon hodoscope setup and online muon event display. (left) The detector with 32-channel preamplifiers, discriminator-shaper, shift-register card and micro-PC readout; (right) the online LabVIEW plot of a muon event.

## 4. Conclusion

A thinner-THGEM with 200 μm thickness has been studied. A 15.9% energy resolution (FWHM) and a gas gain in excess of $10^3$ were obtained at relatively low voltages, especially in Ne mixture and low iC$_4$H$_{10}$ concentration Ar mixture. The experimental results show that the energy resolution depends strongly on the drift electric field. The thinner-THGEM has been used successfully in X-ray diffraction and muon hodoscope experiments.

## Acknowledgments

This work was supported by the National Natural Science Foundation of China (10775181, 10775151) and the State Key Laboratory of Particle Detection and Electronics. We would like to express sincere appreciation to: A. Breskin of the Weizmann Inst. for his stimulating discussions and encouragement; P. Picchi and Rui de Oliveira of CERN for their kind support; and S. Dalla Torre and S. Levorato of Trieste for helpful discussion.